# Digital personal health libraries: a systematic literature review

Huitong Ding[1,2]; Chi Zhang[3]; Ning An[1,2]*; Lingling Zhang[4]; Ning Xie[5,6]; Gil Alterovitz[5,6]

[1]School of Computer Science and Information Engineering, Hefei University of Technology, Hefei, China; [2]Key Laboratory of Knowledge Engineering with Big Data of Ministry of Education, Hefei University of Technology, Hefei, China; [3]Department of Computer Science, University of California Davis, CA, USA; [4]College of Nursing and Health Sciences, University of Massachusetts Boston, Boston, MA, USA; [5]Biomedical Cybernetics Laboratory, Division of General Internal Medicine and Primary Care, Department of Medicine, Brigham and Women's Hospital, Boston, MA, USA; [6]Department of Medicine, Harvard Medical School, Boston, MA, USA

*Corresponding author:

Ning An, PhD

Key Laboratory of Knowledge Engineering with Big Data of Ministry of Education

School of Computer Science and Information Engineering

Hefei University of Technology

Danxia Road No.485, Building-A502

Hefei, Anhui, 230601, China

Phone: +86-180-1995-6086

E-mail: ning.g.an@acm.org.

# Abstract

**Objective:**

This paper gives context on recent literature regarding the development of digital personal health libraries (PHL) and provides insights into the potential application of consumer health informatics in diverse clinical specialties.

**Materials and Methods:**

A systematic literature review was conducted following the Preferred Reporting Items for Systematic Reviews and Meta-Analyses (PRISMA) statement. Here, 2,850 records were retrieved from PubMed and EMBASE in March 2020 using search terms: personal, health, and library. Information related to the health topic, target population, study purpose, library function, data source, data science method, evaluation measure, and status were extracted from each eligible study. In addition, knowledge discovery methods, including co-occurrence analysis and multiple correspondence analysis, were used to explore research trends of PHL.

**Results:**

After screening, this systematic review focused on a dozen articles related to PHL. These encompassed health topics such as infectious diseases, congestive heart failure, electronic prescribing. Data science methods included relational database, information retrieval technology, ontology construction technology. Evaluation measures were heterogeneous regarding PHL functions and settings. At the time of writing, only one of the PHLs described in these articles is available for the public while the others are either prototypes or in the pilot stage.

**Discussion:**

Although PHL researches have used different methods to address problems in diverse health domains, there is a lack of an effective PHL to meet the needs of older adults.

**Conclusion:**

The development of PHLs may create an unprecedented opportunity for promoting the health of older consumers by providing diverse health information.

# INTRODUCTION

## Background and significance

With the continuous enhancement of health awareness, people are more willing to seek health information to manage their health. Health care has begun to shift from the traditional physician-centered paradigm to the patient-centric paradigm.[1] Studies have shown that this new paradigm can bring many benefits, including improved user satisfaction and health outcomes, and reduced health care costs.[2-4] As a representative of this new paradigm, consumer health informatics applications, such as online health communities and decision support systems, enhance the engagement of patients in their health management.[5,6] The recent advancement in data science, including information retrieval and storage technology, enables people to obtain and manage a wide range of complex information about multiple health domains. Health-related information could come from diverse sources, including scientific literature, social media, websites, clinical trials, and community resources. The information sources are increasingly digital and dynamic, and are full of new data types. The wealth of information and the advancement of Internet services offer the potential to revolutionize the paradigm of health decisions and management. However, as shown in Brown et al. (2010),[7] seeking health information online is not as easy as people expect, especially for older adults. Therefore, there is an urgent need for a tool to help people to cope with the challenges in making a valid selection and use information relevant to them.

Digital personal health library (PHL) has increasingly become a popular consumer health tool for the collection, storage, management, and sharing of health information in many forms. Besides integrating information from different sources, PHL also enables consumers to assess multiple functions of the library with a visual and user-friendly interface. With this revolutionary technology, patients expressed increased interest in their health management. Especially in the recent outbreak of COVID-19, the shelter-in-place requirement significantly affects the traditional way of obtaining health information, which takes healthcare providers as the primary information source. Hence, PHLs stand out for their advantage of remote access to information. At the same time, the human factors, including patient engagement and social determinants, are valuable elements to be considered in the successful construction of a PHL. The ability to access information on health care, disease prevention, and health promotion diminishes with age.[8, 9] The global population aging not only creates a vast customer base of older adults but also puts forward new requirements for PHLs. To meet users' different needs for PHL services, researchers need to develop novel approaches to address users' limited health literacy and to improve usability.

When the Internet was just beginning to take off at the opening of this century, Gunther Eysenbach et al[10] summarized in detail the barriers to consumers' access to health information, which set the standard for future research. In recent years, interdisciplinary researcher teams have made great efforts to promote health self-management and to advance in consumer health informatics.[11] By summarizing the current development of PHL, this paper explores whether previous work has sufficiently addressed the initial barriers, what problems still exist after 20 years of development, more importantly, identifies the direction of future development.

# Objective

This paper is devoted to describe and assess the following aspects of PHL studies: (1) health topic and target population; (2) study purpose; (3) library function; (4) data source and scientific method; and (5) evaluation measure and status. It further discusses the trust-building between users and PHLs, and the emergence of novel technologies for promoting patient engagement. Finally, it proposes recommendations on leveraging state-of-the-art technology to develop PHL that helps individuals gather, manage, and use data and information for their health.

# METHODS

This paper conducted a review based on the 2009 PRISMA statement.[12]

# Search strategy and data sources

This paper searched PubMed and EMBASE to retrieve articles published up to March 2020. Search terms were selected based on the concept of a PHL. The principle of the search strategy is to include as many articles as possible and maintain reasonable retrieval accuracy. Therefore, the search items for database queries are "Personal" AND "Health" AND "Library". As shown in Figure 1, searches returned 1,114 articles from PubMed and 2,372 articles from EMBASE, of which 636 were duplicates.

# Inclusion and exclusion criteria

The inclusion criteria for articles in this review were as follows: (1) the article contained the information about authors, title, abstract, and publication year; (2) the article was written in English; and (3) the abstract of the article mentioned health library, a medical health setting, data science method and use of the PHL for health information access, the management or other similar purposes. Literature reviews and duplicate articles were excluded from the analysis. Search results containing the above necessary information were exported from each database.

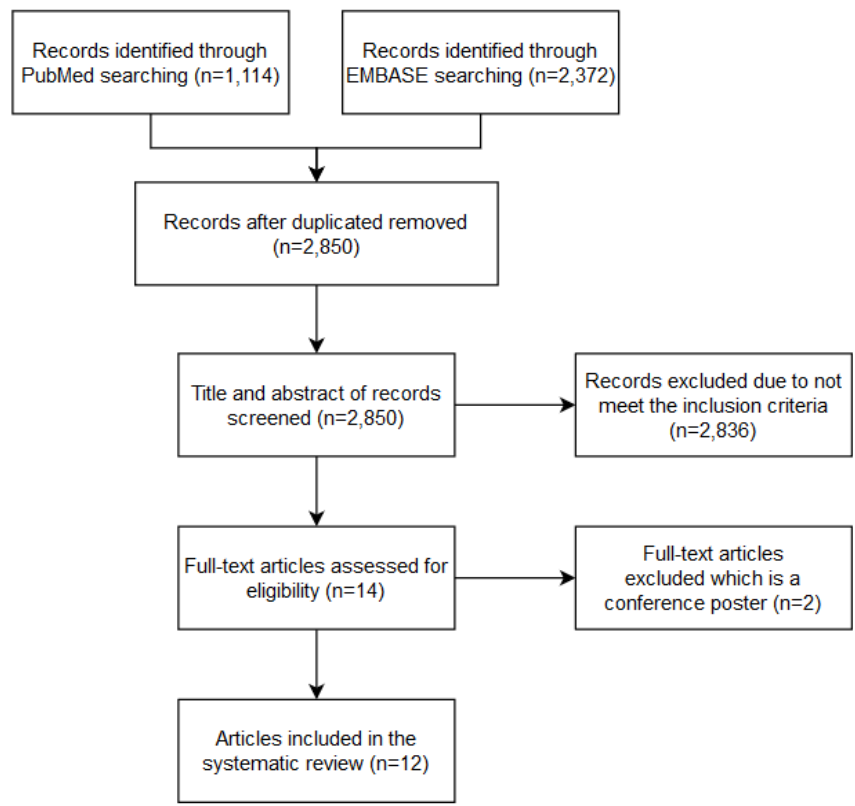

**Figure 1.** PRISMA flow diagram of the article screen process.

Two authors (H. D, C. Z) independently screened the title and abstract of each retrieved article. With the inclusion criteria, the relevance of an article is marked as "yes", "no" or "maybe". Inconsistent annotations and articles labeled with "maybe" were fully discussed to come to an agreement. For complex cases where consensus was difficult to reach, a third author joined the discussion and achieved the final result based on majority voting. For the articles identified as relevant, H. D and C. Z further independently reviewed the full text of them. The disagreements were settled through discussions.

## Data extraction and knowledge mining

The included studies were analyzed from eight key aspects: health topics, target population, study purpose, library function, data sources, data science methods, status, and evaluation measures. This paper did not conduct a quality assessment, as there are no general evaluation criteria for PHL studies to the best of our knowledge.

To better describe the emerging field in PHLs, it is essential to capture the relationships among a core set of keywords using mathematical and statistical methods in Bibliometrics[13]. Accordingly, this paper used the keyword co-occurrence network[14] with the Kamada-Kawai layout[15] to mine and visualize the high-level knowledge structure in PHL literature. This paper also performed multiple correspondence analysis (MCA)[16] to draw a conceptual structure of the field. MCA performed a homogeneity analysis of a Document x Word matrix to obtain the representation of the terms in a low-dimensional Euclidean space.[17] The K-means method was further adopted to identify clusters of terms that express common concepts. The results are interpreted based on the relative positions of the points. The more similar the distribution of the terms, the closer the points they represent in the graph.[18] This paper conducted these knowledge mining analysis based on the R package Bibliometrix[19].

## RESULTS

This review included twelve articles upon screening. The majority of eligible articles were indexed in EMBASE (83.3%, n=10), with the other 2 articles (16.7%) indexed in PubMed. Publication years of these articles ranged from 2004 to 2019, with more than 83% of articles published in the past ten years. Table 1-4 provided the eight aspects of content extracted from these articles, indicating great diversity in the PHL studies. The following sections will explore them in detail.

## Health topics and target populations

As shown in Table 1, the current PHL studies covered a wide range of health domains, with a focus on specific diseases, including infectious diseases[20], congestive heart failure[21], motor neuron disease[22], diabetes[23, 24], and schizophrenia spectrum illnesses[25]. Besides, the rest six studies worked on improving other health-related issues, including electronic prescribing[26], clinical practice guidelines (CPGs) in medical care[27], unhealthy behaviors[28], individualized patient care plans[29], co-operative healthcare work[30], and personal health records (PHRs)[31]. The target populations for these PHLs included patient[20-26, 28-31], health professionals[20], clinicians[21, 29], care providers[22, 27], prescribers[26], and healthcare personnel[30]. Few existing studies had targeted for specific age groups except for the study in[22]. However, this study only indicated that the age of the target population was the typical age of onset of the disease, and did not specify a certain age range. A handful of studies in the paper described the age range of participants they hired to conduct pilot studies, including the studies in[23-25].

**Table 1.** List of health topics and target populations.

| Study | Health Topic | Target Population |
| --- | --- | --- |
| Jay A. Brown, 2010[20] | Infectious diseases | All people especially for health professionals; no age range was specified |
| S. Marceglia et al., 2015[21] | Home monitoring for congestive heart failure patients | Congestive heart failure patients, clinicians; no age range was specified |
| Rose Maunsell et al., 2019[22] | Motor neuron disease (MND) | MND patients and their caregivers; the age range was broadly representative of the MND population |
| Stuart J. Nelson et al., 2009[26] | Electronic prescribing | Prescribers and patients; no age range was specified |
| Dácil Alvarado-Martel et al., 2015[23] | Type 1 diabetes mellitus (T1DM) | T1DM patients; median age was 34 (range 18–50) |
| Robert Moskovitch et al., 2004[27] | CPGs in medical care | Care providers; no age range was specified |
| Sean A. Kidd et al., 2019[25] | Schizophrenia spectrum illnesses | Schizophrenia patients; mean age was 31.4 (range 19–61) |
| Steven H. Woolf et al., 2006[28] | Unhealthy behaviors | Patients; no age range was specified |
| Dina Demner-Fushman et al., 2013[29] | Individualized patient care plans | Clinicians and patients; no age range was specified |
| Marco Masseroli et al., 2006[30] | Co-operative healthcare work | Healthcare personnel including the medical administrator, technicians, physicians, nurses, and patients; no age range was specified |
| Maaike C.M. Ronda et al., 2015[24] | Diabetes care | Type1 and type2 diabetes patients, mean age was 59.7 |
| Jin Sun et al., 2018[31] | Personal Health Records (PHR) | Patients; no age range was specified |

# Study purposes

The study purposes of included articles had strong ties with their health topics and target populations (Table 2). The PHLs' main objectives were to support a better quality of care[21, 27, 29], facilitate patients' self-management[25], support patients' decision-making[22], promote co-operative healthcare work[30], and collect, classify, index, share and provide valuable health information[20, 28, 31]. In general, the current PHLs served a variety of purposes.

**Table 2.** Summary of the study purposes and library functions of all eligible studies.

| Study | Study Purpose | Library Function |
| --- | --- | --- |
| Jay A. Brown, 2010[20] | To collect, classify, index and store information about 275 infectious diseases for continuous refining and updating | Assist medical and public health professionals in actively participating in the surveillance cycle and diagnose infectious diseases earlier and more accurately |
| S. Marceglia et al., 2015[21] | To improve the quality of care by proposing a standards-based information exchange architecture between mHealth Apps and electronic health records | Allow the bi-directional data exchange between patients and healthcare professionals |
| Rose Maunsell et al., 2019[22] | To support U. K. patients making the gastrostomy feeding decision | Provide evidence-based information on gastrostomy placement; communicating the risks and benefits; enable patients to make a decision congruent with their values and appropriate for them |
| Stuart J. Nelson et al., 2009[26] | To encourage an early positive experience of e-prescribing of prescribers by reducing some adoption barriers | Look up medications; create, refill, assess, save, and print a prescription |
| Dácil Alvarado-Martel et al., 2015[23] | To facilitate self-management of T1DM patients by developing a virtual platform | Offer a chat room, calculate body mass index, count carbohydrate, calculate insulin-dose, document on prevention and treatment of acute complications, nutrition, exercise, and more |
| Robert Moskovitch et al., 2004[27] | To improve the adoption and integration of CPGs at the point of care, as part of the evidence-based medicine approach | Converse, search, retrieve, and browse multiple CPG-specification representations (ontologies) |
| Sean A. Kidd et al., 2019[25] | To facilitate illness self-management for schizophrenia patients | Address social isolation through activity scheduling, personalized prompts, and others; foster engagement in the recovery process through evidence-informed contents; facilitate strategy/tip-sharing between A4i users by a peer-peer engagement platform; daily wellness and goal attainment check-ins to inform content delivery and highlight mental health trajectories; passively collected data on phone use as a proxy for sleep and activity levels |
| Steven H. Woolf et al., 2006[28] | To provide beneficial behavior change information | Collect behavioral histories and assess stages of readiness to change by an intake assessment; provide an individually tailored resource library; provide links to local resources; and tailor health advice and printouts for clinicians |
| Dina Demner-Fushman et al., 2013[29] | To provide individualized evidence for care plan | Personalize information, access evidence sources, and provide a details-on-demand display |
| Marco Masseroli et al., 2006[30] | To support co-operative work and share patient information securely among healthcare personnel | Ubiquitous collection, organized storage, fast and easy retrieval of patient data; real-time clinical monitor of patients' states; support the primary physician and nurse activities |
| Maaike C.M. Ronda et al., 2015[24] | To get insight into patients' experiences with a web library called 'Digitaal Logboek' | Access clinic notes and other of medical consultations information; support high-quality diabetes care by providing general diabetes information and an overview of all examinations and visits; allow home measured glucose levels upload and contact with personal care |
| | | provider by secured e-messaging |
| Jin Sun et al., 2018[31] | To enable patients to store and share PHRs on a cloud server safely | Reduce the computational and storage burden by developing a searchable encryption scheme based on cloud-fog computing |

# Library functions

The library functions met the health needs of a specific target population. In addition to the essential information retrieval and browsing functions, the PHLs were able to integrate and utilize information, engaging various forms of health decision support[20, 32], information sharing, and communication among patients and health personnel[23, 32]. However, there was a lack of mining and visualizing deep-level knowledge from online information.

# Data sources and science methods

Different data science methods were used to construct PHLs with multiple functions (Table 3). In addition to web development techniques, the primary approaches included relational database[20], distributed database technology, mHealth App technology[21], data security[26], standards-based information exchange[26], information retrieval technology, ontology construction technology, human-computer interaction[27], cross-platform software development technology[25], information system technology[29], the attribute-based encryption technology and search encryption technology[31]. Most articles did not provide technical details.

**Table 3.** List of data sources and data science methods among the included studies.

| Study | Data Sources | Data Science Methods |
|---|---|---|
| Jay A. Brown, 2010[20] | Textbooks, journal articles, online resources, and electronic databases | Relational database |
| S. Marceglia et al., 2015[21] | OpenMRS, an open-source longitudinal EHR | OpenMRS forms, mHealth App technology |
| Rose Maunsell et al., 2019[22] | Literature reviews, cross-sectional, semi-structured interviews with patients, caregivers and HCP participants | A validated model for web-based DAs |
| Stuart J. Nelson et al., 2009[26] | Personal medication records maintained by patients via MyMedicationList | Data security, standards-based information exchange technology |
| Dácil Alvarado-Martel et al., 2015[23] | Baseline assessment from patients | Database and web technology |
| Robert Moskovitch et al., 2004[27] | Source guidelines downloaded from websites, and Marked up guidelines | Information retrieval technology, ontology construction technology, human-computer interaction |
| Sean A. Kidd et al., 2019[25] | User feedback | Substitutable Medical Applications and Reusable Technologies (SMART) App launch framework, web API development technology based on the Model-View-Controller framework, cross-platform software development technology |
| Steven H. Woolf et al., 2006[28] | Self-reported behavioral history, Web pages from national organizations and agencies, local resources including Web services offered by patients' practices | Web technology |
| Dina Demner-Fushman et al., 2013[29] | National Institutes of Health Clinical Center EHR, Sunrise Enterprise™ 5.5 | Information system technology |
| Marco Masseroli et al., 2006[30] | Collected from different healthcare sites | Web-based systems, Internet technology, distributed database technology |
| Maaike C.M. Ronda et al., 2015[24] | Information from medical consultations, electronic health records | Web-based systems, Internet technology |
| Jin Sun et al., 2018[31] | PHR | The attribute-based encryption technology and search encryption technology |

## Evaluation measures and statuses

Except for one study that did not evaluate the performance of its PHL[20], all the others indicated that their PHLs showed promising results in promoting patient's self-management, although valid assessments were not uniformly performed (Table 4). Around 33.3% (n=4) of studies adopted user satisfaction as the primary evaluation measure. However, the measures varied even for the PHLs targeting the same disease like T1DM. Dácil Alvarado-Martel et al[23] recruited 29 T1DM patients to use the platform. After six months, a set of online questionnaires, including Audit of Diabetes-Dependent Quality of Life[33], Diabetes Treatment Satisfaction Questionnaire (DTSQ)[34], and 12-item Well-Being Questionnaire[35] were used to assess user satisfaction on the efficacy of the virtual platform. Focusing on T1DM patients as well, Maaike C.M. Ronda et al use several validated questionnaires such as DTSQ[34], Problem Areas in Diabetes[36, 37], Diabetes Management Self-Efficacy Scale[38], and Brief Diabetes Knowledge Test[39], to evaluate the following aspects respectively: satisfaction with diabetes treatment, diabetes-specific distress, self-efficacy and diabetes knowledge.

Pre-post comparison at intervention and control practice was another critical way to examine whether exposure to such a PHL improved health outcome. Steven H. Woolf et al[28] collected patients' data through an email-based questionnaire when they have used the PHL for 1 and 4 months. Then the study assessed the health behavior changes of patients from the baseline. However, the evaluation required a large quantity of user data to be statistically valid. How to encourage PHL usage is a factor to be considered to solve this problem.

Notably, approximately 25% of the studies did not report the number of distinct patients from which feedback was obtained, and only one-third reported patient demographic characteristics. Studies assessing PHLs' outcomes often did not list the exact contents of the surveys or interviews; instead, many stated that all evaluations were in the form of a questionnaire, limiting the ability to compare and summarize results across studies.

**Table 4.** Evaluation measures and status of PHLs from all included articles.

| Study | Evaluation Measures | Status |
| --- | --- | --- |
| Jay A. Brown, 2010[20] | None | Could develop an Internet interface within a few months |
| S. Marceglia et al., 2015[21] | Proof-of-concept | Prototype |
| Rose Maunsell et al., 2019[22] | Users' needs, attitudes, and perceived benefits | Pilot study |
| Stuart J. Nelson et al., 2009[26] | Has not yet been thoroughly evaluated | Prototype |
| Dácil Alvarado-Martel et al., 2015[23] | User satisfaction collected by online questionnaire | Still ongoing |
| Robert Moskovitch et al., 2004[27] | Formal functional evaluation | Pilot study |
| Sean A. Kidd et al., 2019[25] | App use metrics by semi-structured interview | Ready for clinical trial and validation testing |
| Steven H. Woolf et al., 2006[28] | Pre-post comparison at intervention and control practices, data on health behaviors, readiness to change, and user satisfaction | Publicly available |
| Dina Demner-Fushman et al., 2013[29] | Usability evaluation with emphasis on the usefulness of the providing information; a focus group discussion; continuous online survey; monthly analysis of system logs | Deployed |
| Marco Masseroli et al., 2006[30] | Clinical pilot projects | Prototype |
| Maaike C.M. Ronda et al., 2015[24] | Questionnaire for usability and satisfaction evaluation | Deployed |
| Jin Sun et al., 2018[31] | Simulation experiments in the cloud-fog environment | Scheme |

# Knowledge discovery

This section presents other mined knowledge in the following figures. As shown in Figure 2, JAMIA is the largest source of references beyond JMIR, BMJ BRIT MED J. This shows the long-term contribution that JAMIA has made to the development of PHLs.

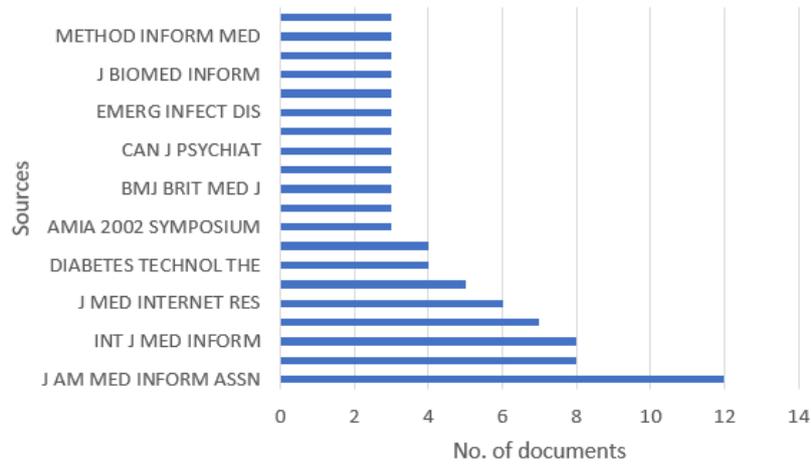

**Figure 2.** Top ten sources of references in the 12 articles.

Figure 3 displays a keyword co-occurrence network. The nodes represent the keywords plus of the article, and the edges between the nodes indicate that the corresponding keywords appear in the same article. The thicker the edges are, the higher the number of articles with the corresponding two keywords. Web of Science produces keywords plus for publications that allow researchers to retrieve a broader range of relevant information.[40] They are keywords extracted from the titles of papers cited in each new article in the database in the Institute for Scientific Information. Many studies have shown that keywords plus describe the content in more detail than the author-assigned keywords.[41, 42] As seen from figure 3, patient and technology are the key nodes connecting keywords in the PHL research field.

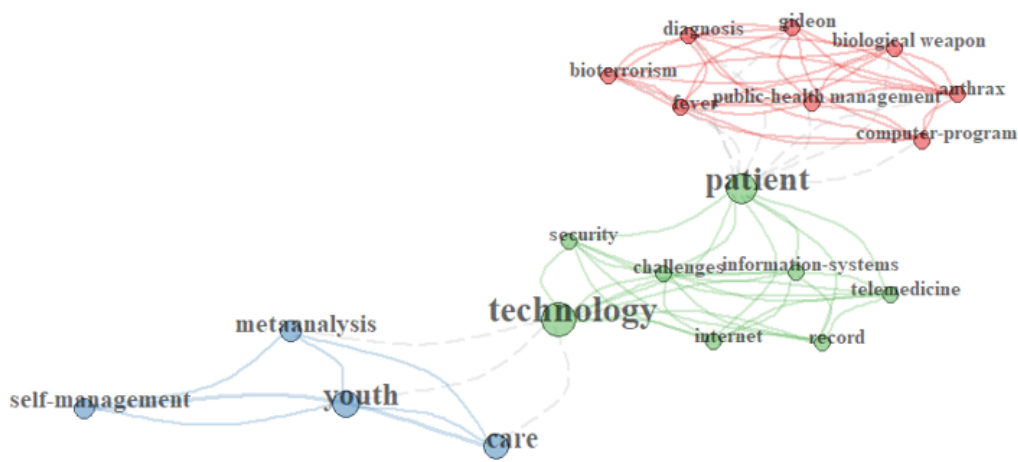

**Figure 3.** Keyword co-occurrence network.

Figure 4 is a two-dimensional conceptual structure map. It showed that the most common title terms were grouped into three clusters, corresponding to the three aspects of consumer health informatics. Forming the concept of consumer, cluster 1 (blue) bought together terms such as patients, diabetes. Articles used these terms had similar research themes in improving patients' outcomes[43, 44] and behavior[28]. Cluster 2 (green) focused on informatics concepts, such as a web-based decision support system. Cluster 3 (red) mainly referred to general health concepts, which were in all articles.

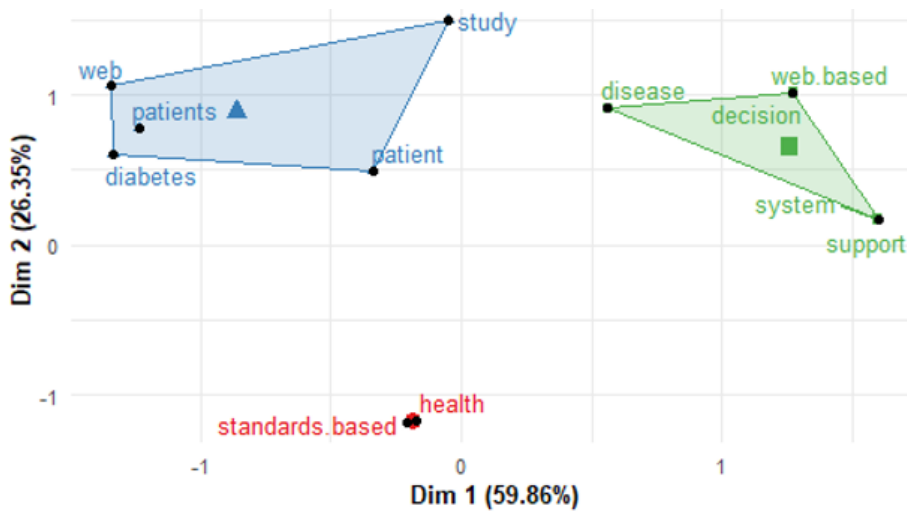

**Figure 4.** Conceptual structure map of the terms of the title.

*The variance explained by the first axis (Dim 1) is 59.86%, and by the second axis (Dim 2) is 26.35%.*

# DISCUSSION

This paper screened 2,850 articles retrieved from PubMed and EMBASE databases to get a general picture of the latest progress on PHL researches, identify gaps, and put forward the prospects of further study. It considered diverse topics addressed in each article, and selected eight aspects to help describe the diversity of the research with comprehensive information for comparison. Overall, the review analysis found that existing PHL research tended to (1) use data science method to enhance self-management of diseases, (2) provide platforms and information resources to improve the efficiency of healthcare, and (3) provide convenient services for various health domains. After 20 years of development, information accessibility has been continuously increasing, including the generation of many professional databases available to the public, and enhanced access to network services. This development eases barriers to access quality information. However, consumers' access to information still faces barriers such as low health literacy and information quality control.

The existing few studies are insufficient to meet the soaring demand for PHL, especially in light of the COVID-19 outbreak. Notably, there are few personalized tools for older people to collect and use cognitive health information. Older adults often find it challenging to handle effective information technology or knowledge to meet their needs because of the complexity of these technologies and a lack of training in tool use.[45] To a certain extent, this has dampened the enthusiasm for the development of PHLs for older adults. Although some studies have designed user-friendly interfaces, researchers did not pay enough attention to health literacy. Health literacy represents skills, knowledge, and the expectations that health professionals have of the public's interest in and understanding of health information and services.[46] Research has shown that people's limited health literacy[47] dramatically affects their ability to use health information and constrains the popularization of PHL. More research should study the association between effective use of PHL and health literacy. The development of face-to-face instructions and user-friendly guides will be the next important step in supporting older adults to learn and use PHL in their daily lives effectively.

User trust is critical in the development of consumer health informatics.[48] The lack of trust in online privacy and security affects the popularity of PHLs. Consumers need to provide highly sensitive personal information, including medical conditions, to use the services of PHLs. In particular, with the trend of personalized service, users must provide information, including social determinants, so that the PHLs can deliver information accurately through a recommendation algorithm. Another part of the distrust stems from the quality of the information provided on the PHLs. Users may wonder whether the information provided on the PHL for disease intervention and medication is correct. PHLs should recognize the importance of consumer trust and take steps to maintain it, for example, publishing online privacy policies.[49] Recently, researchers began to improve the privacy and data security of websites on a technical level. Jin Sun et al. proposed a security scheme that guarantees safe storage and sharing of patients' PHRs on a cloud server while keeping personal information confidential.[31] However, this framework has not yet been deployed in practice.

The social determinants of health (SDoH) have gradually gained recognition as a way to address health disparities.[50] The researchers and doctors defined SDoH as the "conditions in which people are born, grow, live, work, and age that shape health".[51] It has been shown to have a significant influence on health outcomes.[52] Society and government agencies have begun to take measures to address this problem actively.[53] However, according to the findings of this paper, SDOH has not been well addressed in the existing PHLs when providing information services to users.

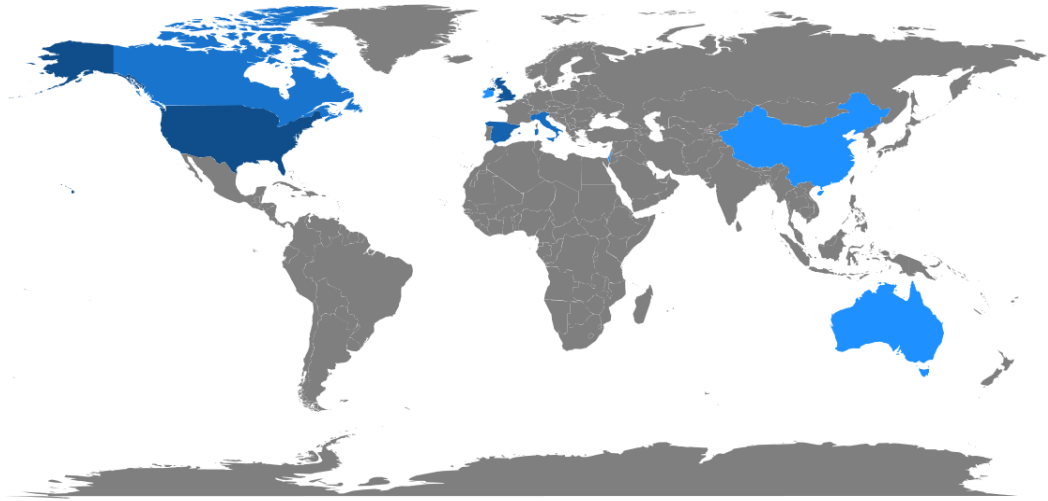

**Figure 5.** Country scientific production in the PHL domain in the scope of English paper.

A particular technology or application of PHL is not expected to be available to all soon.[54] There is still a digital divide[55] between the developed and the developing countries in PHL research, even as technologies such as the Internet seem to become more affordable. Figure 5 indicates that the number of research production of developed countries is higher than that of developing countries. It takes time for the PHL to become widely accessible.

**Future directions**

Based on the above findings, this paper summarized several aspects of PHL worthy of further exploration, including the following topics.

**Improve information quality**

Quality evaluation of health information from the Internet has always been critical for PHL and other applications in consumer health informatics. However, the threshold of online information dissemination has been lowered along with the emergence of new online media platforms and the popularity of social media. This change has led to many sites containing unverified information. In particular, people with lower health literacy have weaker discrimination. Therefore, future studies should focus on developing an effective data quality assessment method to verify the quality of information sources and ensure the high quality of the underlying data of knowledge engineering.

**Develop knowledge mining and recommendation algorithms**

Data-driven models can derive insights from health information by interpreting and extracting valuable patterns from online data. These modeling techniques include mathematical models and various machine learning methods, especially knowledge graphs and recommendation algorithms. At present, the PHLs focus on the collection, retrieval, and display of information but lack the further processing of massive information. Research has shown that graphical display has many advantages, including promoting healthy behaviors and reducing errors caused by narratives.[56, 57] Knowledge graph provides a way to mine, analyze, construct, and display knowledge and their interconnections. It has gradually been applied to the field of health informatics, from electronic medical records[58] to health risk prediction[59]. Using this technology to improve the effectiveness of PHL is a promising development direction.

After mining out high-quality health information, the recommendation algorithms can help deliver this information to the most interested user. Data science models combining with user's information (age, gender, browsing behavior, and more) may help the user to obtain accurate, personalized health information.

**Integrate multiple functions**

PHLs varied widely in medical applications and user populations, including the specific functions that they provide, the outcomes evaluated, and the data sources, making it more challenging to conduct a comparable statement on the libraries' practicability for self-management. Besides, it can force people to register multiple PHLs in order to obtain information about different diseases, which complicates the information acquisition process. At the same time, personal data is stored on multiple websites, which increases the possibility of privacy disclosure and puts forward higher requirements for data security. The integration of multiple functions not only solves the above problems but also collects more comprehensive user behavior data, enhancing the learning process of PHL's recommendation algorithm, and delivering more accurate information for users. Future research could explore how individuals and various combinations of PHL elements support disease management and prevention mechanisms.

**Standardize evaluation method**

While many researchers employed similar outcome measurements, such as effectiveness, efficiency, and satisfaction, the methods for gathering those measurements varied widely. Different evaluation methods prevented a direct comparison of libraries' performance. Therefore, efforts are urgently needed to standardize the evaluation method for identifying the challenges of improving the quality of PHL.

**Consider social determinants comprehensively**

With the rapid expansion of the PHL, differences in user health outcomes will likely continue to persist with differences in social determinants such as technology access.[60-62] Income level, Internet access, or other living conditions, may play a role in PHL engagement. For example, although older adults have a greater need for online health information, their inadequate health literacy makes them the most difficult demographic group to use the PHL. Therefore, collecting users' social determinants and further exploring their influence on user interaction will improve personalized service and make the PHL more engaging. The PHL should be made more attractive to many older people, women, and minorities. This attraction lies in the design, language, and culture issues contained and addressed in the PHL development. The causes of the digital divide are also an important topic to explore. As researchers focus on improving the technological infrastructure, they should be aware that public attitudes may also be contributing to the gap. Although technological limitations such as Internet access ability may be a factor restricting the development of PHL in developing countries, it is more important to change public attitude regarding technology, taking it as a necessity to achieve personal health. This paper is to improve people's awareness of health information acquisition in regions where technology is not too backward.

**Limitations**

Although this article attempted to minimize the possibility of missing articles by generalizing search terms, retrieval of PubMed and EMBASE did not necessarily yield all the relevant articles. Also, due to the substantial heterogeneity of the PHLs covered by the reviewed articles, reliable, quantitative, and comprehensive evaluation of the research results were limited. Finally, the screen of articles partly influenced by the authors' prior knowledge and understanding of the field, which was relatively subjective.

# CONCLUSION

PHL has found a way into a variety of health professional applications, including disease management, healthcare, unhealthy behaviors, and electronic prescribing. However, there is still a great need for PHL that caters to the elderly. By examining the details of these PHL studies in eight aspects from health topics to evaluation measures, this systematic review identified the deficiencies in current work and outlined the priorities for future development. Also, this paper adopts the knowledge mining method to discover that patient and technology are the key factors connecting this field. PHL would benefit a lot from improving information quality, applying knowledge mining and recommend algorithms, integrating function, standardizing evaluation method, and thoroughly taking into account social determinants.

# FUNDING

This work was partially supported by the National Key R&D Program of China (No. 2018YFB1003204), Anhui Provincial Key Technologies R&D Program (No. 1804b06020378), and the National Institutes of Health under award number 7R01GM118467-05.

# CONTRIBUTORS

All authors contributed significantly to this work by participating in the following task: study design, article collection, publication screening, information extraction, data analysis, manuscript writing. All authors reviewed and approved this manuscript before submission.

**Conflict of interest statement**

None to be declared.